\documentclass[sigconf, screen]{acmart}

\usepackage{tcolorbox}
\usepackage{fbox}
\usepackage{xcolor}
\usepackage{booktabs}
\usepackage{verbatim}
\usepackage{pgfplots}
\usepackage{filecontents}
\usepackage{tikz}
\usepackage{subfigure}
\usepackage{graphics}
\usepackage{multirow}
\usepackage{booktabs}
\usepackage{graphicx}
\usepackage{float}
\usepackage{tabularx}
\usepackage{array}
\usepackage{enumitem}
\usepackage[all]{nowidow}
\usepackage[multiple]{footmisc}
\usepackage[flushleft]{threeparttable}
\usepackage{enumitem}
\setlist{leftmargin=4mm}
\usetikzlibrary{patterns}
\usepackage{balance}

\pgfplotsset{compat=1.16}

\definecolor{red}{rgb}{1, 0, 0}
\definecolor{green}{rgb}{0, 1, 0}
\definecolor{blue}{rgb}{0, 0, 1}

\newcolumntype{L}[1]{>{\raggedright\arraybackslash}p{#1}}
\newcolumntype{C}[1]{>{\centering\arraybackslash}p{#1}}

\newcommand{\rv}[1]{#1}
\newcommand{\nrv}[1]{#1}

\AtBeginDocument{%
  \providecommand\BibTeX{{%
    \normalfont B\kern-0.5em{\scshape i\kern-0.25em b}\kern-0.8em\TeX}}}

\copyrightyear{2022} 
\acmYear{2022} 
\setcopyright{acmcopyright}\acmConference[ICSE '22]{44th International Conference on Software Engineering}{May 21--29, 2022}{Pittsburgh, PA, USA}
\acmBooktitle{44th International Conference on Software Engineering (ICSE '22), May 21--29, 2022, Pittsburgh, PA, USA}
\acmPrice{15.00}
\acmDOI{10.1145/3510003.3510155}
\acmISBN{978-1-4503-9221-1/22/05}

\begin{document}

\title[Log-based Anomaly Detection with Deep Learning: How Far Are We?]{Log-based Anomaly Detection with Deep Learning: \\ How Far Are We?}


\author{Van-Hoang Le} 
\affiliation{%
  \institution{The University of Newcastle}
  \state{NSW}
  \country{Australia}
}
\email{vanhoang.le@uon.edu.au}

\author{Hongyu Zhang}
\affiliation{%
  \institution{The University of Newcastle}
  \state{NSW}
  \country{Australia}
}
\email{hongyu.zhang@newcastle.edu.au}
\authornote{Hongyu Zhang is the corresponding author.}

\begin{abstract}
Software-intensive systems produce logs for troubleshooting purposes. Recently, many deep learning models have been proposed to automatically detect system anomalies based on log data. These models typically claim very high detection accuracy. For example, most models report an F-measure greater than 0.9 on the commonly-used HDFS dataset. To achieve a profound understanding of how far we are from solving the problem of log-based anomaly detection, in this paper, we conduct an in-depth analysis of five state-of-the-art deep learning-based models for detecting system anomalies on four public log datasets. Our experiments focus on several aspects of model evaluation, including training data selection, data grouping, class distribution, data noise, and early detection ability. Our results point out that all these aspects have significant impact on the evaluation, and that all the studied models do not always work well. The problem of log-based anomaly detection has not been solved yet. Based on our findings, we also suggest possible future work.
\end{abstract}



\maketitle

\vspace{-1mm}
\section{Introduction}
\label{sec:introduction}
High availability and reliability are essential for large-scale software-intensive systems. As these systems provide various services to a large number of users, a small problem in the system could lead to user dissatisfaction and even significant financial loss. Anomaly detection is, therefore, important for the quality assurance of complex software-intensive systems.

\begin{figure}[h]
    \centering
    \includegraphics[width=0.96\linewidth]{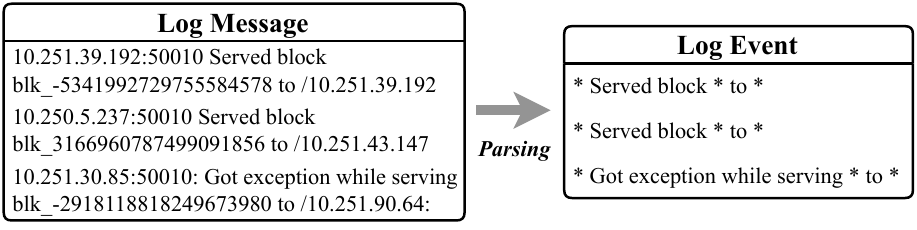}
    \vspace{-6pt}
    \caption{An example of log messages and log events}
    \label{fig:log_snippet}
    \vspace{-2pt}
\end{figure}

Software-intensive systems often record runtime information by printing console logs. 
A large and complex system could produce a massive amount of logs, which can be used for troubleshooting purposes. For example, the cloud computing systems of Alibaba Inc. produce about 30-50 gigabytes (around 120-200 million lines) of tracing logs per hour~\cite{mi2013toward}.
Log data is usually unstructured text messages, which can help engineers understand the system's internal status 
and facilitate monitoring, administering, and troubleshooting of the system~\cite{he2020survey}. Log messages can be parsed into log events, which are templates (constant part) of the messages. \rv{Figure \ref{fig:log_snippet} shows an example of raw log messages and the corresponding log events obtained after parsing.}

Over the years, many data-driven methods have been proposed to automatically detect system anomalies by analyzing log data~\cite{breier2015anomaly, zhang2020anomaly, du2017deeplog, chen2004failure, xu2009detecting, zhang2019robust,logcluster}. For example, He et al.~\cite{he2016experience} evaluated six popular machine learning (ML) algorithms for log-based anomaly detection. These ML-based methods share some limitations of inflexible features, inefficiency, and weak adaptability~\cite{zhang2019robust}. In order to address these issues, deep learning (DL) has been adopted and produced promising results. Du et al.~\cite{du2017deeplog} proposed to use Long-Short Term Memory (LSTM) to model the sequential patterns of normal sessions, then identify anomalies as those violated the patterns. Meng et al.~\cite{meng2019loganomaly} trained an LSTM model to detect sequential and quantitative anomalies using log count vectors as inputs. They also proposed {template2vec} to consider the synonyms and antonyms of the words in log templates. Other studies~\cite{zhang2019robust, li2020_swisslog} represent log templates as semantic vectors to handle the instability of log data. Generally, the existing DL-based log anomaly detection methods show promising results on commonly used datasets and claim their superiority over traditional ML-based approaches. For instance, DeepLog~\cite{du2017deeplog} and LogAnomaly~\cite{meng2019loganomaly} all reported a very good performance on commonly-used HDFS and BGL datasets (with F-measure values greater than 0.9).

However, we notice that several important aspects are overlooked by the existing work. These aspects are associated with experimental datasets, evaluation metrics, 
and experimental settings. In this work, we would like to dive deep into the problem and answer: \textit{Are log-based anomaly detection methods with deep learning as good as they claimed? What are the major factors that could affect their performance?}

To answer the above questions, we conduct a systematic evaluation of five representative deep learning models for log-based anomaly detection (including DeepLog~\cite{du2017deeplog}, LogAnomaly~\cite{meng2019loganomaly}, PLELog~\cite{yang2021semi}, LogRobust~\cite{zhang2019robust}, and CNN~\cite{lu2018detecting}) on four datasets (including HDFS, BGL, Spirit, and Thunderbird), under controlled experimental settings. 
We first conduct an analysis of training data selection and grouping techniques. Then we explore the impact of different characteristics of datasets (including data noise and class distribution) on model performance. Finally, we analyze the ability of the models in the early detection of anomalies. 
Through extensive experiments, we obtain the following major findings about the current deep learning models for log-based anomaly detection:
\begin{itemize}
\item The training data selection strategies (random or chronological) have significant impact on the semi-supervised log-based anomaly detection models. Randomly selecting training data could cause the data leakage problem and unreasonably high detection accuracy.

\item Different log data grouping methods have substantial influence on the performance of the models. Models tend to lose their accuracy when dealing with shorter log sequences. 

\item The effectiveness of the models is significantly affected by the highly imbalanced class distribution. Commonly-used metrics, including Precision, Recall, and F-measure, are not comprehensive enough for evaluating a log-based anomaly detection model with highly imbalanced data. 

\item A small amount of data noise, including mislabelled logs and log parsing errors can downgrade anomaly detection performance. Compared to semi-supervised methods, supervised models are more sensitive to mislabelled logs in the training data. 
Models capable of understanding the semantic meaning of log data could reduce the impact of log parsing errors.

\item Different models have different abilities in the early detection of system anomalies. Some models can detect anomalies earlier than others.

\end{itemize}

\noindent In summary, the major contributions of this work are as follows: 
\begin{itemize}
\item We conduct an extensive evaluation of five representative deep learning models for log-based anomaly detection. 

\item We conclude that the existing models are not evaluated comprehensively and do not generalize well in different experimental settings. 

\item Based on the evaluation results, we point out the advantages and disadvantages of existing models, and suggest future research work for log-based anomaly detection. 
\end{itemize}


\vspace{-1mm}
\section{Log-based Anomaly Detection with Deep Learning}
\label{sec:background}

\subsection{Representative Models}
\label{sec:RepresentativeModels}
In recent years, many deep learning-based models have been proposed to analyze log data and detect anomalies~\cite{breier2015anomaly, zhang2020anomaly, du2017deeplog, zhang2019robust, nedelkoski2020self, yang2021semi, li2020_swisslog}. Some of these models use supervised learning techniques (such as LogRobust~\cite{zhang2019robust} and CNN~\cite{lu2018detecting}), while others employ semi-supervised approaches (such as DeepLog~\cite{du2017deeplog}, LogAnomaly~\cite{meng2019loganomaly}) or unsupervised approaches ~\cite{farzad2020unsupervised}.
Some recent representative models are as follows:

\textbf{DeepLog}. Du et al.~\cite{du2017deeplog} proposed to utilize an LSTM model to learn the system's normal executions by predicting the next log event given preceding events. It detects anomalies by determining whether or not an incoming log event violates the prediction results of the LSTM model.
Their experimental results show that DeepLog can achieve an F-measure of 0.96 on the HDFS dataset. 

\textbf{LogAnomaly}. Meng et al.~\cite{meng2019loganomaly} proposed LogAnomaly, which uses log count vectors as inputs to train an LSTM model. They also proposed \textit{template2vec}, a synonyms and antonyms based method, to represent log templates as semantic vectors to match new log events with existing templates. 
Like DeepLog, a forecasting-based detection model is designed to predict the next log event, and if the examined log event violates the prediction results, it will be marked as an anomaly. LogAnomaly can achieve F-measures of 0.95 and 0.96 on HDFS and BGL datasets, respectively.

\textbf{PLELog}. Yang et al.~\cite{yang2021semi} addressed the issue of insufficient labels via probabilistic label estimation and designed an attention-based GRU neural network to detect anomalies. The GRU-based detection model is built to classify log sequences into two classes, normal or abnormal.
Their experimental results indicate that PLELog outperforms existing semi-supervised methods and achieves high performance on HDFS and BGL datasets (i.e., 0.96 and 0.98, respectively).

\textbf{LogRobust}. Zhang et al.~\cite{zhang2019robust} incorporated a pre-trained Word2vec model, namely FastText~\cite{joulin2016fasttext}, and combined it with TF-IDF weights to learn the representation vectors of log templates. Then, these vectors were input to an Attention-based Bi-LSTM model to detect anomalies. The experimental results show that LogRobust can address the instability of log events and achieve F-measures of 0.99 on the original HDFS dataset and 0.89-0.96  
on synthetic datasets.

\textbf{CNN}. Lu et al.~\cite{lu2018detecting} applied a Convolutional Neural Network (CNN) for log-based anomaly detection. Logs are grouped into sessions, then transformed into a trainable matrix. A CNN model is trained using this matrix as inputs to classify a log sequence into normal or abnormal. The CNN model can achieve an F-measure of 0.98 on the HDFS dataset.

In this study, we systematically evaluate the above five models. DeepLog and LogAnomaly adopt a forecasting-based approach (i.e., detecting anomalies by predicting the next log event given preceding log sequences), while PLELog, LogRobust, and CNN are classification-based models (i.e., detecting anomalies by classifying log sequences).
Apart from these models, there are some other deep learning-based approaches. 
For example, \textit{Logsy}~\cite{nedelkoski2020self} utilizes the Transformer network~\cite{vaswani2017attention} to detect anomalies from log data. \textit{AutoEncoder} has been employed in~\cite{farzad2020unsupervised} to detect log anomalies in an unsupervised manner along with Isolation Forest.
\textit{SwissLog}~\cite{li2020_swisslog} proposes to use a dictionary-based log parser and an Attention-based Bi-LSTM network to detect anomalies for diverse faults.
As their source code is not publicly available, we do not experimentally evaluate these models in this study.

\vspace{-6pt}
\subsection{The Common Workflow}
\begin{figure}[h]
    \centering
    \includegraphics[width=.98\linewidth, height=7.6cm]{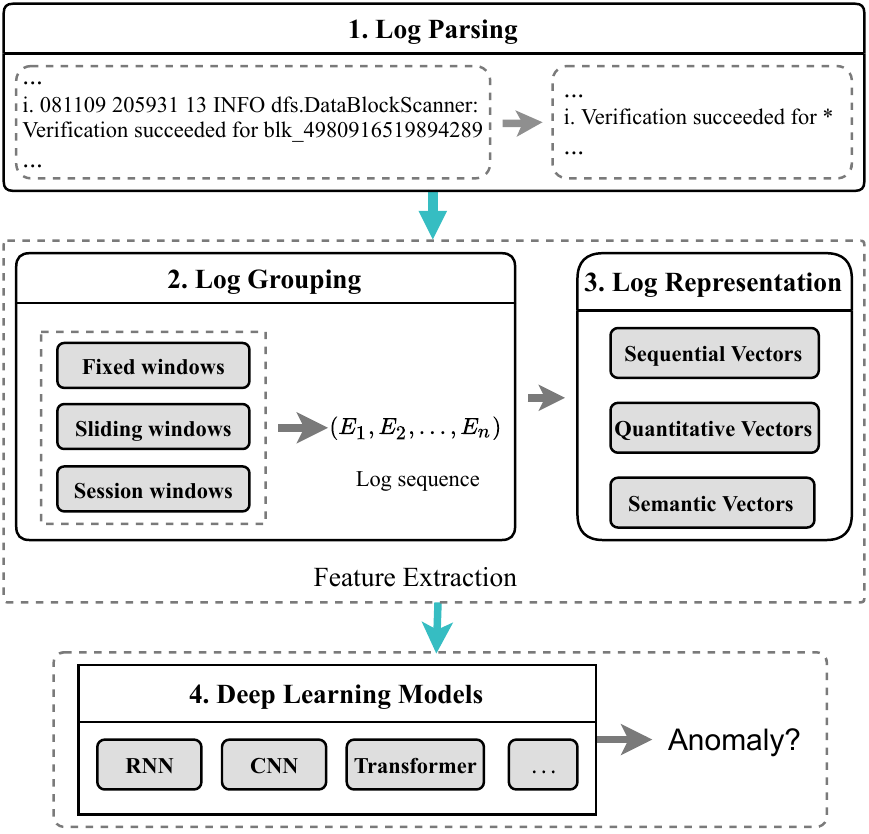}
    \vspace{-6pt}
    \caption{Log-based Anomaly Detection Workflow with Deep Learning: The Common Workflow}
    \label{fig:overflow}
\end{figure}

The common overall framework of DL models for log-based anomaly detection is shown in Figure \ref{fig:overflow}. Generally, the framework consists of four steps: (1) log parsing, (2) log grouping, (3) log representation, (4) anomaly detection through DL models.

\vspace{-3pt}
\subsubsection{Log parsing}
\label{sec:log_parsing}
Logs are semi-structured texts, which contain various fields such as timestamp and severity. To favor downstream tasks, log parsing is applied to automatically convert each log message into a specific event template (constant part) associated with parameters (i.e., variable part).
For example, the log template \textit{``Verification succeeded for $*$"} can be extracted from the log message \textit{``Verification succeeded for blk\_4980916519894289"} in Figure \ref{fig:overflow}. Here, \textit{``$*$"} denotes the position of a parameter.

There are many log parsing techniques, based on frequent pattern mining~\cite{nagappan2010abstracting, vaarandi2015logcluster, dai2020logram}, clustering~\cite{tang2011logsig, hamooni2016logmine, shima2016length}, and heuristics~\cite{he2017drain, jiang2008abstracting, makanju2009clustering}. The heuristics-based approaches make use of the characteristics of logs and have been found to perform better than other techniques in terms of accuracy and time efficiency~\cite{zhu2019tools}.

\vspace{-3pt}
\subsubsection{Log Grouping}
The main purpose of this step is to separate logs into various groups, where each group represents a finite chunk of logs~\cite{he2016experience}. These groups are called \textit{log sequences}, from which features are extracted and fed into anomaly detection models. As introduced in~\cite{he2016experience}, three types of windows are applied (see Figure \ref{fig:overflow}) for log grouping, including:
\textit{(1) Fixed window.} Logs are grouped by fixed windows according to their occurrences.  
The occurrence is defined by the timestamp of log messages or by the order of its appearance in the dataset. Each window has a fixed size (i.e., window size), which means the time span or the number of logs.
\textit{(2) Sliding window.} Sliding windows consist of two attributes: window size and step size. The window size can be the time span or the number of logs in a log sequence, while the step size is the forwarding distance. 
\textit{(3) Session window.} Different from fixed/sliding windows, session windows are based on the identifier of logs. Identifiers are used to group logs in the same execution path. For example, HDFS logs use \textit{block\_id} to record the execution path. 


\vspace{-3pt}
\subsubsection{Log Representation}
After log grouping, logs are represented in different formats 
required by DL models. 
Existing DL-based anomaly detection models convert logs into three main types: (1) sequential vectors, (2) quantitative vectors, and (3) semantic vectors. Sequential vectors reflect the order of log events in a window. For example, DeepLog~\cite{du2017deeplog} assigns each log event with an index, then generates a sequential vector for each log window. Quantitative vectors are similar to log count vectors, which are used to hold the occurrence of each log event in a log window. LogAnomaly~\cite{meng2019loganomaly} leverages both sequential and quantitative vectors to detect anomalies. Different from them, semantic vectors are acquired from language models to represent the semantic meaning of log events. Each log window is converted into a set of semantic vectors for the detection models. For instance, LogRobust~\cite{zhang2019robust} adopts a pre-trained FastText~\cite{joulin2016fasttext} model to compute the semantic vectors of log events.

\vspace{-3pt}
\subsubsection{Deep Learning Models}
\label{sec:detection_models}
After the log representation phase, the extracted features are fed to deep learning models for the anomaly detection task. 
A variety of DL techniques have been applied to log-based anomaly detection:
1)~\textbf{RNN}. Recurrent Neural Networks (RNNs), including its variants Long Short-Term Memory (LSTM)~\cite{hochreiter1997long} and Gated Recurrent Units (GRUs)~\cite{cho2014learning}, are neural networks designed to handle the sequential inputs with arbitrary length. 
Bi-directional RNN is used to represent the sequential text in both directions (i.e., forward and backward). 
RNN and its variants have been applied in many studies on log-based anomaly detection. Specifically, DeepLog~\cite{du2017deeplog} and LogAnomaly ~\cite{meng2019loganomaly} use LSTM to predict the next log event. LogRobust~\cite{zhang2019robust} applies an Attention-based Bi-LSTM network to handle the instability of logs, while PLELog adopts GRUs to build a classification model.
2) \textbf{CNN}. A Convolutional Neural Network (CNN) uses convolution operation to extract meaningful local patterns of input.
In ~\cite{lu2018detecting}, a CNN model originally designed for sentence classification~\cite{kim-2014-convolutional} is applied to log-based anomaly detection.
3) \textbf{Transformer}. The Transformer network has made significant progress in neural machine translation and related pretraining tasks in recent years.
It has been applied in~\cite{nedelkoski2020self,le2021log,guo2021logbert} for log-based anomaly detection.

\vspace{-1mm}
\section{Study Design}
\label{sec:study_design}
\subsection{Motivation}
\label{sec:motivation}

Recent studies~\cite{meng2019loganomaly, li2020_swisslog} have shown that deep learning-based approaches can achieve high accuracy (e.g., F-Measure higher than 90\%) on commonly used datasets (e.g., the HDFS dataset). These results seem to suggest that the problem of anomaly detection can be solved almost perfectly through deep learning. 
To 
explore the actual effectiveness of existing DL models for log-based anomaly detection ($models$ for short in the rest of the paper), 
we would like to evaluate the models from the following aspects \nrv{associated with the common workflow of the models:}

\vspace{-3pt}
\subsubsection{The selection of training data}
The results of anomaly detection could be affected by the selection of training data. 
\nrv{In some studies (such as DeepLog~\cite{du2017deeplog}, LogRobust~\cite{zhang2019robust}, PLELog~\cite{yang2021semi}, and CNN~\cite{lu2018detecting}), the training and test data are selected based on the timestamp of logs. We call it $chronological$ strategy.
Other studies~\cite{guo2021logbert, chen2021experience, meng2019loganomaly} apply fixed/sliding windows to group log events into log sequences, then shuffle all logs sequences before splitting them into training and testing sets.  We call it $random$ strategy. 
The random strategy allows models to see more log events and achieve high accuracy (e.g., higher than 88\% F-measure for all models~\cite{chen2021experience}).} However, this strategy may lead to the data leakage problem in the training phase. That is, it is possible that the training set contains parts of future data, and the testing set contains parts of past data, thus making it not suitable for real-world scenarios where we only use historical logs to build a detection model. 
The effectiveness of the models with different training data selection strategies (random or chronological) should be investigated. 

\vspace{-3pt}
\subsubsection{The grouping of log data}
The log data can be grouped into sequences by session, sliding, or fixed windows. 
Choosing a proper window size is challenging. For example, if the window size is small, the models face difficulty in capturing those anomalies that span multiple sequences. On the other hand, if the window size is large, log sequences might include multiple anomalies and confuse the detection scheme~\cite{liu2021lognads}. In this work, we evaluate \rv{
many window sizes, including the window sizes from 20 to 200 log messages as well as the window sizes of 0.5 hour and 1 hour}, and also 
session window~\cite{he2016experience, zhang2020anomaly}. 

\vspace{-3pt}
\subsubsection{The imbalanced class distribution}
\label{sec:imbalanced_data}

In literature, much research work~\cite{zhang2008distribution, zhang2009investigation, fenton2000quantitative} has shown that in a large software system, the distribution of faults is skewed - that a small number of modules accounts for a large proportion of the faults. In our work, we find that the distribution of anomalies is also skewed. 
The anomalous log sequences usually account for the minority of the dataset, which can be only from 0.5\% to 15\% of a dataset, as shown in Section \ref{sec:datasets}. The highly imbalanced data imposes challenges for anomaly detection. In general, it is difficult for a machine learning technique to identify a small number of anomalies from a large amount of logs. 
The imbalance between normal and abnormal classes could cause the model to perform poorly.

The highly imbalanced class distribution also has implications for evaluation metrics.
The performance of log-based anomaly detection models is usually measured by Precision, Recall, and F-Measure ~\cite{du2017deeplog, zhang2019robust, guo2021logbert, meng2019loganomaly, he2016experience}. However, previous studies~\cite{zhang2007comments} pointed out that prediction results may not always be satisfactory in the presence of imbalanced data distribution. In this work, we would like to explore if the commonly-used metrics can effectively evaluate the effectiveness of a log-based anomaly detection model under the scenario of imbalanced class distribution.

\vspace{-3pt}
\subsubsection{The quality of data}  

For the evaluation of log-based anomaly detection models, labeled data is required. The commonly-used public datasets (such as HDFS and BGL) are manually inspected and labeled by engineers. Data noise (false positives/negatives) may be introduced during the manual labeling process. 
Although the data noises only occupy a small portion of logs, they could downgrade the performance of existing models.
The noise can be from the errors in the preprocessing phase (i.e., log parsing).
The logging statements could also frequently change during software evolution, resulting in new log events that were not seen in the training phase~\cite{zhang2019robust, meng2019loganomaly}. Zhang et al.~\cite{zhang2019robust} found that 30.3\% of logs are changed in the latest version based on their empirical study on Microsoft online service systems. Le et al.~\cite{le2021log} found that log parsing errors can lead to many incorrect log events, thus downgrading anomaly detection performance. 
Therefore, we would like to investigate the effectiveness of the models with different degrees of data noise.

\vspace{-3pt}
\subsubsection{Early detection ability}
\label{sec:motivation_leadtime}
System anomalies can affect the normal operations of upper-layer software applications and significantly affect users' experience. If no actions are taken, more severe problems or even service interruptions may occur. Therefore, it is important that the anomalies are captured earlier, so that more mitigation actions could be taken. An effective anomaly detection model should be able to identify the early signals of system anomalies, detect the anomalies as early as possible, and meanwhile achieve high detection accuracy. This is especially essential for the online detection scenario, where anomalies are detected on the fly.


\vspace{3pt}
Because of the above concerns, we argue that the capabilities of deep learning-based techniques for anomaly detection should be re-evaluated. In this work, we design experiments to measure the impact of these factors on five representative DL models for log-based anomaly detection.

\vspace{-6pt}
\subsection{Evaluated Models}
In this study, we evaluate the five representative models described in Section \ref{sec:RepresentativeModels}, namely DeepLog, LogAnomaly, PLELog, LogRobust, and CNN. These models have their source code publicly available, and we can confirm the correctness of source codes by reproducing results presented in their original paper.
Specifically, we adopt the public implementations~\cite{wuyifan2021deeplog, donglee2021logdeep} of DeepLog and LogAnomaly.
For LogAnomaly, the template2vec model is trained with domain-specific antonyms and synonyms adding by operators. Since this information is unavailable, we use a pre-trained FastText word2vec mode~\cite{joulin2016fasttext} to compute the semantic vectors of log templates. The template vector is calculated as the weighted average of the vectors of the template's words. 
For LogRobust, we adopt the implementation provided by its authors and convert it into a PyTorch-based implementation. For PLELog, we leverage its public implementation~\cite{plelog2021}.
For models whose hyperparameter settings are reported in their paper, we use the same hyperparameter values. Otherwise, we tune their hyperparameters empirically.

\vspace{-6pt}
\subsection{Datasets}
\label{sec:datasets}
	
To evaluate the studied models for log-based anomaly detection, we select four public datasets~\cite{loghub2021, he2020loghub}, namely HDFS, BGL, Thunderbird, and Spirit. The details of each dataset are as follows:
\begin{itemize}
    \item \textbf{HDFS (Hadoop Distributed File System)} dataset is produced from more than 200 Amazon EC2 nodes. In total, the HDFS dataset consists of 11,175,629 log messages. These log messages form different log windows according to their \textit{block\_id}, reflecting a program execution in the HDFS system. There are 16,838 blocks of logs (2.93\%) in this dataset indicating system anomalies.
    \item \textbf{BGL (Blue Gene/L)} dataset is a supercomputing system log dataset collected by Lawrence Livermore National Labs (LLNL) \cite{oliner2007supercomputers}. The dataset contains 4,747,963 log messages. Each message in the BGL dataset was manually labelled as either normal or anomalous. There are 348,460 log messages (7.34\%) that were labelled as anomalous.
    \item\textbf{Spirit} dataset is an aggregation of system log data from the Spirit supercomputing system at Sandia National Labs \cite{oliner2007supercomputers}. There are more than 172 million log messages labeled as anomalous on the Spirit dataset. In this paper, we use a small set containing the first 5 million log lines of the original Spirit dataset, which contains 764,500 abnormal log messages (15.29\%).
    \item \textbf{Thunderbird} dataset is an open dataset of logs collected from a Thunderbird supercomputer at Sandia National Labs (SNL) \cite{oliner2007supercomputers}. The log data contains normal and abnormal messages which are manually identified. Thunderbird is a large dataset of more than 200 million log messages. We leverage 10 million continuous log lines for computation-time purposes, which contain 4,934 abnormal log messages (0.49\%).
\end{itemize}

\noindent Table \ref{tab:overview_datasets} summarizes the statistics of datasets used in our experiments.

\begin{table*}[h!]
\caption{The statistics of datasets used in the experiments}
\vspace{-6pt}

\resizebox{.93\textwidth}{!}{
\setlength{\tabcolsep}{1.5pt}
\renewcommand{\arraystretch}{1.12}
\begin{tabular}{ccccccccc} 
\toprule
\multirow{2}{*}{\begin{tabular}[c]{@{}c@{}}\\\textbf{Dataset}\end{tabular}} & \multirow{2}{*}{\textbf{\# Log Events}} & \multirow{2}{*}{\textbf{Grouping}} & \multirow{2}{*}{\textbf{\# Log sequences}} & \multirow{2}{*}{\begin{tabular}[c]{@{}c@{}}\textbf{\# Avg. seq.}\\\textbf{length}\end{tabular}} & \multicolumn{2}{c}{\textbf{Training Data}}                        & \multicolumn{2}{c}{\textbf{Testing Data}}                          \\ 
\cline{6-9}
                                                                            &                                         &                                    &                                            &                                                                                                 & \textbf{\textit{\# Log sequences}} & \textbf{\textit{\# Anomaly}} & \textbf{\textit{\# Log sequences}} & \textbf{\textit{\# Anomaly}}  \\ 
\hline
HDFS                                                                        & 48                                      & session (random)                   & 575,061                                    & 19.4                                                                                            & 460,048                            & 13,521 (2.9\%)               & 115,013                            & 3,317 (2.9\%)                 \\ 
\hline
\multirow{4}{*}{BGL}                                                        & \multirow{4}{*}{1,847}                  & 1 hour (random)                    & 3,606                                      & 1,307.1                                                                                         & 2,884                              & 536 (18.6\%)                 & 722                                & 129 (17.9\%)                  \\ 
\cline{3-9}
                                                                            &                                         & 1 hour (chron.)                    & 3,606                                      & 1,307.1                                                                                         & 2,625                              & 496 (18.9\%)                 & 981                                & 171 (17.4\%)                  \\ 
\cline{3-9}
                                                                            &                                         & 100 logs (chron.)                  & 47,135                                     & 100                                                                                             & 37,708                             & 4,009 (10.6\%)               & 9,427                              & 817 (8.7\%)                   \\ 
\cline{3-9}
                                                                            &                                         & session (random)                   & 69,252                                     & 68.1                                                                                            & 55,401                             & 25,066 (45.2\%)              & 13,851                             & 6,309 (45.5\%)                \\ 
\hline
\multirow{3}{*}{Spirit}                                                     & \multirow{3}{*}{2,880}                  & 1 hour (random)                    & 1,173                                      & 4,262.6                                                                                         & 1,001                              & 882 (88.11\%)                & 172                                & 71 (41.28\%)                  \\ 
\cline{3-9}
                                                                            &                                         & 1 hour (chron.)                    & 1,173                                      & 4,279.0                                                                                         & 938                                & 760 (81.02\%)                & 235                                & 192 (81.7\%)                  \\ 
\cline{3-9}
                                                                            &                                         & 100 logs (chron.)                  & 50,000                                     & 100                                                                                             & 40,000                             & 19,384 (48.5\%)              & 10,000                             & 346 (3.5\%)                   \\ 
\hline
\multirow{3}{*}{Thunderbird}                                                & \multirow{3}{*}{4,992}                  & 1 hour (random)                    & 209                                        & 47,651.5                                                                                        & 167                                & 42 (25.1\%)                  & 42                                 & 6 (14.29\%)                   \\ 
\cline{3-9}
                                                                            &                                         & 1 hour (chron.)                    & 209                                        & 47,651.5                                                                                        & 169                                & 40 (23.7\%)                  & 40                                 & 8 (20.0\%)                    \\ 
\cline{3-9}
                                                                            &                                         & 100 logs (chron.)                  & 99,593                                     & 100                                                                                             & 79,674                             & 816 (1.0\%)                  & 19,919                             & 27 (0.1\%)                    \\ 
\bottomrule
\hline
\multicolumn{4}{l}{%
  \begin{minipage}{.5\linewidth}%
    Note: chron. denotes the chronological strategy.%
  \end{minipage}%
}\\
\end{tabular}
}
\label{tab:overview_datasets}
\vspace{-6pt}
\end{table*}



\vspace{-6pt}
\subsection{Research Questions}
The goal of this study is to analyze the performance of the representative deep learning models for log-based anomaly detection models. 
\rv{We design the following research questions in accordance with the evaluation aspects described in Section~\ref{sec:motivation}}.


\vspace{1pt}
\textbf{RQ1: How do the existing approaches perform with different training data selection strategies?}
We want to evaluate whether or not the studied models are able to achieve good accuracy with different training data selection strategies.
To this end, we conduct experiments with two different strategies for training data selection:
\textit{(1) Random selection}: For each dataset, we first sort logs by timestamps, and then apply the fixed window grouping technique to generate log sequences. Next, these log sequences are shuffled, and split into training/testing sets with the ratio of 80:20.
\textit{(2) Chronological selection}: For each dataset, we utilize the first 80\% of raw logs (that appear in chronological order) for training and the remaining 20\% for testing. Next, we apply the fixed window grouping technique to generate log sequences. \rv{We do not shuffle the generated log sequences in this strategy.} Therefore, we can guarantee that only historical logs are used in the training phase, and there are no future logs used in this phase.
    
In this RQ, we experiment on the BGL, Thunderbird, and Spirit datasets. The window size for fixed window grouping is set to 1 hour. As the HDFS dataset does not contain timestamp information, the chronological selection cannot be applied to HDFS, and thus it is not used in this RQ.

\vspace{1pt}
\textbf{RQ2: How do the existing models perform with different data grouping methods?}
To evaluate the impact of different data grouping methods on the performance of anomaly detection models, we choose the following three data grouping methods, including: (1) fixed-window grouping with the window size of 1 hour (as used in RQ1) \rv{and 0.5 hour}, (2) fixed-window grouping with \rv{the window size varying from 20 to 200 log messages}, and (3) session window grouping. For the first case, we use BGL, Spirit, and Thunderbird datasets. For the case of session windows, \rv{we use \textit{block\_id} and \textit{node\_id} to group logs on HDFS and BGL datasets, respectively.}

\vspace{1pt}
\textbf{RQ3: Can the existing approaches work with different class distributions?}
As shown in Section \ref{sec:datasets}, our subject datasets \nrv{represent highly imbalanced class distributions with anomaly ratios that can be only 0.1\% (i.e., on the Thunderbird dataset)}. To perform a more systematic evaluation, we simulate different imbalanced scenarios by randomly removing the normal/abnormal log sequences from the \rv{subject datasets}. For a real-world production system, the number of anomalies is much less than the number of normal events~\cite{lin2018predicting, zhang2008distribution, zhang2009investigation}. Therefore, we vary the imbalance ratio from 0.1\% to 15\%, which indicates the percentage of anomalies in the dataset. In this way, we create six synthetic datasets with the imbalance ratio of 0.1\%, 0.5\%, 1\%, 5\%, 10\%, and 15\%.

\vspace{1pt}
\textbf{RQ4: Can existing approaches work with different degrees of data noise?}
To evaluate the impact of mislabelled logs on the performance of the studied models, we randomly add some anomalies (from 1\% to 10\%) into the training data for semi-supervised methods. For supervised methods, we randomly change the label of a specific portion (from 1\% to 10\%) to simulate the mislabelled logs. In this way, we create five synthetic datasets with the mislabelled proportion of 1\%, 2\%, 5\%, 8\%, and 10\%. 
Moreover, to measure the impact of noises from log parsing errors, we experiment with four commonly used log parsers, including Drain \cite{he2017drain}, Spell \cite{du2016spell}, AEL \cite{jiang2008abstracting}, and IPLoM \cite{makanju2009clustering}.

\vspace{1pt}
\textbf{RQ5: How early can the existing models detect anomalies in online detection?}
As described in Section \ref{sec:motivation_leadtime}, a model should not only detect anomalies precisely but also should be able to detect anomalies as early as possible so that more mitigation actions could be taken. Therefore, in this RQ, we evaluate the studied models on four datasets to investigate their ability in early detection of anomalies. To this end, we record the number of examined log messages before each model raises an anomaly alert for a log sequence, in the online detection setting.


\vspace{-6pt}
\subsection{Experimental Setup}
In our experiments, we preprocess the log data and 
conduct DL-based anomaly detection as follows:

\textbf{(1) Log Parsing.}  To extract log templates from log data, we use the log parser Drain~\cite{he2017drain} with the default parameter settings~\cite{logparser2021}. 
The log data is denoted by $L = \{l_1, l_2, ..., l_i, ..., l_{N_L}\}$ and contains $N_L$ entries (i.e., lines) of log messages. \nrv{Each log message $l_i$ is parsed into a log template $E(l_i)$, which is denoted $e_i$ for short.}

\textbf{(2) Log Grouping.} We apply session window to group logs in the HDFS dataset using \textit{block\_id}. Each session is labeled using ground truth. For other datasets (BGL, Spirit, and Thunderbird), we use the fixed window strategy to group log data into $N_S$ chunks (i.e., log sequences), denoted as $S = \{s_1, s_2, ..., s_u, ..., s_{N_S}\}$, \nrv{where $s_u = \{e_i, e_{i+1}, ..., e_j\}$ is a set of log templates}.
According to the ground truth (labeled by domain engineers), a log sequence is abnormal if it contains an anomalous log message according to the ground truth (labeled by domain engineers).
$F$ denotes the size of each log sequence. In this study, \nrv{we vary the value of $F$ from 20, 100, 200 log messages, to 0.5 and 1 hour depending on each research question.}

\textbf{(3) Log Representation.} Log sequences are now converted into numerical vectors, which can be input to a DL model. DeepLog transforms log sequences into sequential vectors 
by assigning each log event with an index. LogAnomaly leverages both sequential vectors and quantitative vectors to train the model. The occurrence of each log event is counted and forms the quantitative vectors, which represent the system execution behaviors~\cite{meng2019loganomaly}. PLELog extracts the semantic vectors of log templates by using a pre-trained Glove model~\cite{pennington2014glove}. Similarly, LogRobust and CNN also convert log sequences into semantic vectors using a pre-trained word2vec model. We adopt the pre-trained FastText~\cite{joulin2016fasttext} model to compute the semantic vectors for LogRobust and CNN. 


\textbf{(4) Deep Learning Model.} In this step, depending on the method, a DL model is trained using the corresponding feature vectors generated from the previous phase. DeepLog and LogAnomaly have two LSTM layers with 128 neurons. LogRobust contains a two-layer Bi-LSTM with 128 neurons and an attention layer. PLELog utilizes a one-layer GRU network. CNN has three Convolutional layers with different filters and a max-pooling layer for feature extraction.

\rv{To avoid bias from randomness, we perform each experiment five times and report the average results.}
We conduct our experiments on a Windows Server 2012 R2 with Intel Xeon E5-2609 CPU, 128GB RAM, and an NVIDIA Tesla K40c. 
\vspace{-3pt}
\subsection{Evaluation Metrics}
To measure the effectiveness of models in anomaly detection, we use the Precision, Recall, Specificity, and F1-Score metrics, which are defined as follows:

\begin{itemize}
    \item \textit{Precision:} the percentage of correctly detected abnormal log sequences amongst all detected abnormal log sequences by the model. $Prec = \frac{TP}{TP + FP}$.
    \item \textit{Recall:} the percentage of log sequences that are correctly identified as anomalies over all real anomalies. $Rec = \frac{TP}{TP + FN}$.
    \item \textit{Specificity}: the percentage of log sequences that are correctly identified as normal over all real normal sequences.\\$Spec=\frac{TN}{TN + FP}$. 
    \item \textit{F-Measure:} the harmonic mean of \textit{Precision} and \textit{Recall}.\\$F1 = \frac{2 * Prec * Rec}{Prec + Rec}$.
\end{itemize}
TP (True Positive) is the number of abnormal log sequences the are correctly detected by the model. FP (False Positive) is the number of normal log sequences that are wrongly identified as anomalies. FN (False Negative) is the number of abnormal log sequences that are not detected by the model.


\vspace{-1mm}
\section{Results and Findings}
\label{sec:evaluation}
\subsection{RQ1: Performance with different training data selection strategies?}
\label{sec:eval_training_data}

For RQ1, we apply fixed window grouping with the size of 1 hour to generate log sequences on BGL, Spirit, and Thunderbird datasets (see Table \ref{tab:overview_datasets}).
The training data is selected by random or chronological selection.
The experimental results are shown in Table \ref{tab:RQ1_training_data}.

We find that, for semi-supervised models (i.e., LogAnomaly, DeepLog, and PLELog), the results with random selection are much better than those with chronological selection.
For example, DeepLog achieves an F-measure of 0.927 with the random selection of training data on the BGL dataset. When training and testing sets are separated by the time order (i.e., chronological selection), the F-measure drops to 0.426. The reason is that with random selection, the models can see future log events in the training phase (i.e., data leakage), \nrv{therefore} they can make more accurate predictions. Besides, DeepLog and LogAnomaly train the models using the index of log events (i.e., sequential and quantitative vectors) and ignore the semantic meaning of logs during the training phase. DeepLog marks any new log events as anomalies and produces many false alarms. LogAnomaly can simply match some unseen log events with those appearing in the training phase, but it is not adequate compared to those models that are trained through the semantic understanding of logs.

The supervised models (i.e., LogRobust and CNN) perform much better than the semi-supervised models on both strategies since the models are trained with a large amount of normal and abnormal data. For example, these two models achieve around 0.94 F-measure on the Thunderbird dataset with the chronological setting, while others perform poorly. Another reason for these results is the advantages of semantic vectors used by these models, which can identify the semantically similar log events and also distinguish different log events \cite{zhang2019robust}. Still, \nrv{we can see that in general the results with random selection are  better than those with chronological selection}.

Our experimental results confirm that models perform better with random selection. 
\nrv{The data leakage problem is a reason for the good performance of some DL-based log anomaly detection models (e.g., LogAnomaly~\cite{meng2019loganomaly}, which uses the random strategy in their evaluation)}.
Due to this problem, we suggest that chronological selection should be applied to evaluate the effectiveness of the models in real-world scenarios. Hence, for other RQs, we will apply the chronological selection to group log messages.

\begin{table*}[htbp]
\caption{\rv{Comparison of model performance with random selection and chronological selection of training data}}
\vspace{-6pt}
\label{tab:RQ1_training_data}
\resizebox{.98\linewidth}{!}{%
\renewcommand{\arraystretch}{1.1}
\setlength{\tabcolsep}{2pt}
\begin{tabular}{ccccccccccccc} 
\toprule
\multirow{2}{*}{\textbf{Model}} & \multicolumn{4}{c}{\begin{tabular}[c]{@{}c@{}}BGL\\ (random/chronological selection)\end{tabular}} & \multicolumn{4}{c}{\begin{tabular}[c]{@{}c@{}}Spirit\\ (random/chronological selection)\end{tabular}} & \multicolumn{4}{c}{\begin{tabular}[c]{@{}c@{}}Thunderbird\\ (random/chronological selection)\end{tabular}}  \\ 
\cline{2-13}
                                 & \textit{\textbf{Prec}} & \textit{\textbf{Rec}} & \textit{\textbf{Spec}} & \textit{\textbf{F1}}     & \textit{\textbf{Prec}} & \textit{\textbf{Rec}} & \textit{\textbf{Spec}} & \textit{\textbf{F1}}        & \textit{\textbf{Prec}} & \textit{\textbf{Rec}} & \textit{\textbf{Spec}} & \textit{\textbf{F1}}              \\ 
\hline
DeepLog                          & 0.952/0.270            & 0.903/0.988           & 0.990/0.437            & 0.927/0.426              & 0.867/0.438            & 1.0/1.0               & 0.386/0.090            & 0.929/0.609                 & 0.232/0.200            & 1.0/1.0               & 0.007/0                    & 0.369/0.333                       \\
LogAnomaly                       & 0.961/0.313            & 0.903/0.798           & 0.992/0.551            & 0.931/0.483              & 0.882/0.438            & 1.0/1.0               & 0.456/0.090            & 0.937/0.609                 & 0.234/0.229            & 1.0/1.0               & 0.014/0.156                & 0.371/0.371                       \\
PLELog                           & 0.963/0.702            & 0.935/0.791           & 0.992/0.899              & 0.949/0.744              & 0.956/0.931            & 0.974/0.767           & 0.818/0.690            & 0.965/0.841                 & 0.584/0.250              & 0.344/1.0             & 0.729/0.250              & 0.414/0.400                       \\
LogRobust                        & 0.972/0.994            & 0.984/0.942           & 0.995/0.999            & 0.981/0.967              & 0.994/0.985            & 0.978/0.915           & 0.979/0.990            & 0.986/0.949                 & 0.803/0.900            & 0.921/1.0               & 0.931/0.969            & 0.941/0.947                        \\
CNN                              & 0.994/0.871              & 0.963/0.947           & 0.999/0.970              & 0.978/0.908              & 1.0/0.986              & 1.0/1.0               & 1.0/0.990              & 1.0/0.993                   & 1.0/0.889              & 0.875/1.0             & 1.0/0.969              & 0.933/0.941                       \\
\bottomrule
\end{tabular}
}
\vspace{-6pt}
\end{table*}

\begin{tcolorbox}[left=2pt,right=2pt,top=0pt,bottom=0pt]
\textbf{Summary.} The training data selection strategies have significant impact on the semi-supervised log-based anomaly detection models. Although the random selection strategy leads to better results than the chronological selection strategy, it could cause the data leakage problem and fail to evaluate the effectiveness of the models in real-world scenarios. 
\end{tcolorbox}
\subsection{RQ2: How do the existing models perform with different data grouping methods?}
\label{sec:eval_data_grouping}
\rv{In RQ1, we use the fixed window size of 1-hour logs.} 
In this RQ, we train the models on three datasets using chronological selection with \rv{fixed window grouping of various sizes (i.e., 20 log messages, 100 log messages, 200 log messages, and 0.5-hour logs)}. Table \ref{tab:fixed_window_100logs} shows the results. 

\begin{table}[h]
\centering
\caption{\rv{Results of models with fixed-window grouping of different sizes}}
\vspace{-6pt}
\label{tab:fixed_window_100logs}
\resizebox{\linewidth}{!}{%
\setlength{\tabcolsep}{1.5pt}
\renewcommand{\arraystretch}{1.05}
\begin{tabular}{cccccccccccccc} 
\hline
\multirow{2}{*}{\textbf{Model}} &                      & \multicolumn{4}{c}{\textbf{BGL}}                                                                                & \multicolumn{4}{c}{\textbf{Spirit}}                                                                             & \multicolumn{4}{c}{\textbf{Thunderbird}}                                                                         \\ 
\cline{2-14}
                                 & \multicolumn{1}{l}{} & \textbf{\textit{20l}} & \textbf{\textit{100l}} & \textbf{\textit{200l}} & \textbf{\textit{0.5h}} & \textbf{\textit{20l}} & \textbf{\textit{100l}} & \textbf{\textit{200l}} & \textbf{\textit{0.5h}} & \textbf{\textit{20l}} & \textbf{\textit{100l}} & \textbf{\textit{200l}} & \textbf{\textit{0.5h}}  \\ 
\hline
\multirow{4}{*}{DeepLog}         & P                    & 0.128                     & 0.166                      & 0.192                      & 0.209                      & 0.504                     & 0.500                        & 0.175                      & 0.291                      & 0.004                     & 0.017                      & 0.005                      & 0.162                       \\
                                 & R                    & 0.995                     & 0.988                      & 0.987                      & 0.984                      & 0.776                     & 0.861                      & 0.985                      & 1.0                        & 0.938                     & 0.963                      & 1.0                        & 1.0                         \\
                                 & S                    & 0.539                     & 0.53                       & 0.528                      & 0.481                      & 0.986                     & 0.969                      & 0.747                      & 0.068                      & 0.899                     & 0.922                      & 0.005                      & 0.088                       \\
                                 & F1                   & 0.227                     & 0.285                      & 0.322                      & 0.345                      & 0.611                     & 0.633                      & 0.298                      & 0.450                       & 0.008                     & 0.033                      & 0.010                       & 0.279                       \\ 
\hline
\multirow{4}{*}{LogAnomaly}      & P                    & 0.136                     & 0.176                      & 0.203                      & 0.276                      & 0.498                     & 0.508                      & 0.198                      & 0.330                       & 0.004                     & 0.025                      & 0                          & 0.154                       \\
                                 & P                    & 0.970                     & 0.985                      & 0.985                      & 0.973                        & 0.773                     & 0.870                       & 0.981                      & 1.0                        & 0.938                     & 0.963                      & 1.0                        & 1.0                         \\
                                 & S                    & 0.581                     & 0.562                      & 0.559                      & 0.643                      & 0.986                     & 0.970                       & 0.783                      & 0.225                      & 0.891                     & 0.950                       & 0.005                      & 0.029                       \\
                                 & F1                   & 0.239                     & 0.299                      & 0.336                      & 0.430                       & 0.606                     & 0.642                      & 0.330                       & 0.496                      & 0.008                     & 0.050                       & 0.009                      & 0.267                       \\ 
\hline
\multirow{4}{*}{PLELog}          & P                    & 0.592                     & 0.595                      & 0.862                      & 0.760                       & 0.375                     & 0.371                      & 0.141                      & 0.516                      & 0.429                     & 0.826                      & 0.692                      & 1.0                           \\
                                 & R                    & 0.882                     & 0.880                       & 0.844                      & 0.785                      & 0.286                     & 0.824                      & 0.552                      & 0.663                      & 0.688                     & 0.704                      & 0.360                       & 0.500                           \\
                                 & S                    & 0.958                     & 0.968                      & 0.985                      & 0.965                      & 0.991                     & 0.950                       & 0.816                      & 0.763                      & 1.0                       & 1.0                        & 1.0                        & 1.0                           \\
                                 & F1                   & 0.708                     & 0.710                       & 0.853                      & 0.772                      & 0.325                     & 0.511                      & 0.225                      & 0.581                      & 0.528                     & 0.760                       & 0.474                      & 0.667                           \\ 
\hline
\multirow{4}{*}{LogRobust}       & P                    & 0.616                     & 0.696                      & 0.684                      & 0.819                      & 0.947                     & 0.943                      & 0.751                      & 0.989                      & 0.377                     & 0.318                      & 0.289                      & 0.458                       \\
                                 & R                    & 0.969                     & 0.968                      & 0.963                      & 0.946                      & 0.979                     & 0.954                      & 0.965                      & 0.989                      & 0.876                     & 1.0                        & 0.960                       & 0.917                       \\
                                 & S                    & 0.959                     & 0.960                       & 0.949                      & 0.971                      & 0.999                     & 0.998                      & 0.982                      & 0.996                      & 0.999                     & 0.997                      & 0.994                      & 0.809                       \\
                                 & F1                   & 0.753                     & 0.810                       & 0.800                        & 0.878                      & 0.963                     & 0.947                      & 0.845                      & 0.989                      & 0.531                     & 0.482                      & 0.444                      & 0.611                       \\ 
\hline
\multirow{4}{*}{CNN}             & P                    & 0.634                     & 0.698                      & 0.744                      & 0.837                      & 0.954                     & 0.959                      & 0.961                      & 0.989                      & 0.907                     & 0.900                        & 0.720                       & 1.0                         \\
                                 & R                    & 0.969                     & 0.965                      & 0.965                      & 0.935                      & 0.966                     & 0.948                      & 0.950                       & 0.989                      & 0.813                     & 0.670                       & 0.720                       & 0.667                       \\
                                 & S                    & 0.962                     & 0.96                       & 0.949                      & 0.974                      & 0.999                     & 0.999                      & 0.998                      & 0.996                      & 1.0                       & 1.0                        & 0.999                      & 1.0                         \\
                                 & F1                   & 0.767                     & 0.810                       & 0.840                       & 0.883                      & 0.960                     & 0.953                      & 0.955                      & 0.989                      & 0.857                     & 0.766                      & 0.720                       & 0.800                         \\
\hline
\multicolumn{12}{l}{%
  \begin{minipage}{\linewidth}%
    20l: 20 logs, 100l: 100 logs, 200l: 200 logs, 0.5h: 0.5-hour logs.%
  \end{minipage}%
}\\
\end{tabular}
}
\vspace{-6pt}
\end{table}

\nrv{The results show that different window sizes lead to different performance of detection models.
We can observe that, on the BGL dataset, compared to the results of 1-hour-logs setting in Table \ref{tab:RQ1_training_data}, the performance of the models mostly drops.
For example, on the BGL dataset, the decrease of F1 measure ranges from 4.84\% (PLELog) to 50.31\% (LogAnomaly) when the 100-log-messages setting is used.
On the Spirit dataset, DeepLog and LogAnomaly achieve the best results when using the window size of 100 log messages, while others perform the best with 0.5-hour-logs setting. Similar results can be found on the Thunderbird dataset, where the detection performance is different across different window sizes.}

We next evaluate the performance of models on session window grouping. Logs can be grouped based on the identifiers (i.e., \textit{node\_id} and \textit{block\_id} for BGL and HDFS datasets, respectively) to represent the execution path of a task. We perform anomaly detection after each session ends~\cite{zhang2020anomaly, zhang2019robust}. We do not evaluate this RQ on other datasets (i.e., Spirit and Thunderbird) because they do not have the identifier information, such as \textit{block\_id}, in their log messages. Table~\ref{tab:session_window} shows the results.

\begin{table}[h]
\centering
\caption{Results of models with session grouping on BGL and HDFS datasets}
\vspace{-3pt}
\label{tab:session_window}
\resizebox{.98\linewidth}{!}{%
\setlength{\tabcolsep}{2pt}
\renewcommand{\arraystretch}{1.1}
\begin{tabular}{ccccccccc} 
\toprule
\multirow{2}{*}{\textbf{Model}} & \multicolumn{4}{c}{\textbf{BGL}}                                                               & \multicolumn{4}{c}{\textbf{HDFS}}                                                               \\ 
\cline{2-9}
                                & \textbf{\textit{Prec}} & \textbf{\textit{Rec}} & \textbf{\textit{Spec}} & \textbf{\textit{F1}} & \textbf{\textit{Prec}} & \textbf{\textit{Rec}} & \textbf{\textit{Spec}} & \textbf{\textit{F1}}  \\ 
\hline
DeepLog                         & 0.997                  & 1.0                   & 0.997                  & 0.998                & 0.835                  & 0.994                 & 0.994                  & 0.908                 \\
LogAnomaly                      & 0.997                  & 1.0                   & 0.998                  & 0.999                & 0.886                  & 0.893                 & 0.961                  & 0.966                 \\
PLELog                          & 0.995                  & 0.992                 & 0.996                  & 0.994                & 0.893                  & 0.979                 & 0.996                  & 0.934                 \\
LogRobust                       & 1.0                    & 1.0                   & 1.0                    & 1.0                  & 0.961                  & 1.0                   & 0.989                  & 0.980                 \\
CNN                             & 1.0                    & 1.0                   & 1.0                    & 1.0                  & 0.966                  & 1.0                   & 0.991                  & 0.982                 \\
\bottomrule
\end{tabular}
}
\vspace{-6pt}
\end{table}

It is obvious that the results using session windows on the BGL dataset are better than those using fixed windows. 
Moreover, compared to the results on the BGL dataset in Table \ref{tab:RQ1_training_data}, 
we can find that the performance is improved using the session window. The reason could be that the log events in an execution path exhibit many relations~\cite{zhang2020anomaly, lou2010mining}, which can be captured and utilized for anomaly detection. On the HDFS dataset, all models also achieve good performance (all F-measure values are higher than 0.9).
\begin{tcolorbox}[left=2pt,right=2pt,top=0pt,bottom=0pt]
\textbf{Summary.} The data grouping methods could have significant impact on the log-based anomaly detection models. With fixed window grouping, models \nrv{tend to perform unsteadily when dealing with different window sizes}. Grouping by session windows could lead to better results. 
\end{tcolorbox}

\subsection{RQ3: How do the existing approaches perform with different class distributions?}
\label{sec:eval_data_distribution}

Table \ref{tab:RQ1_training_data} shows the results on the subject datasets, which have different ratios of anomalies. \nrv{For example, we can see that, on the Spirit dataset, DeepLog and Log-Anomaly achieve high Precision, Recall, and F-measure (all higher than 0.86) with random selection. However, the Specificity results are low (less than 0.5), which reveal that the models actually perform poorly: they classify a lot of normal logs as anomalies in this scenario, causing many false alarms. 
This also happens on the Thunderbird dataset when DeepLog and LogAnomaly mark mostly all of the normal logs as anomalies (Specificity $ \approx $ 0).
In this RQ, to perform a more systematic evaluation, we simulate different imbalanced scenarios with the percentage of anomalies increased from 0.1\% to 15\%. We
use the results on HDFS and BGL datasets to explain our findings, as shown in Table \ref{tab:RQ3_imblance_ratio_hdfs}.}




    
From Table \ref{tab:RQ3_imblance_ratio_hdfs}, we find that when the percentage of anomalies increases, the performance of the models is better, \rv{which is indicated by the increase of all four metrics. For example, the scores of LogRobust on BGL dataset are improved by 63.8\%, 10.1\%, 2.6\%, and 38.4\% in Precision, Recall, Specificity, and F-measure, respectively.}
The result shows that it is more difficult to detect anomalies when the dataset is highly imbalanced.

\begin{table}[h]
\centering
\caption{\rv{Results with different class distributions}}
\vspace{-6pt}
\label{tab:RQ3_imblance_ratio_hdfs}
\resizebox{\linewidth}{!}{%
\setlength{\tabcolsep}{1.5pt}
\renewcommand{\arraystretch}{1.1}
\begin{tabular}{cccccccccccccc} 
\hline
\multirow{2}{*}{\textbf{\textbf{Model}}} &    & \multicolumn{6}{c}{\textbf{HDFS}}                                                                                                                  & \multicolumn{6}{c}{\textbf{BGL}}                                                                                                                    \\ 
\cline{2-14}
                                         &    & \textbf{\textit{0.1\%}} & \textbf{\textit{0.5\%}} & \textbf{\textit{1\%}} & \textbf{\textit{5\%}} & \textbf{\textit{10\%}} & \textbf{\textit{15\%}} & \textbf{\textit{0.1\%}} & \textbf{\textit{0.5\%}} & \textbf{\textit{1\%}} & \textbf{\textit{5\%}} & \textbf{\textit{10\%}} & \textbf{\textit{15\%}}  \\ 
\hline
\multirow{4}{*}{DeepLog}                 & P  & 0.942                   & 0.562                   & 0.638                 & 0.879                 & 0.941                  & 0.962                  & 0.124                   & 0.156                   & 0.184                 & 0.237                 & 0.276                  & 0.278                   \\
                                         & R  & 0.485                   & 0.894                   & 0.995                 & 0.997                 & 0.994                  & 0.994                  & 0.987                   & 0.969                   & 0.991                 & 1.000                 & 0.988                  & 0.959                   \\
                                         & S  & 1.000                   & 0.997                   & 0.994                 & 0.993                 & 0.994                  & 0.994                  & 0.424                   & 0.424                   & 0.417                 & 0.407                 & 0.435                  & 0.440                   \\
                                         & F1 & 0.640                   & 0.690                   & 0.777                 & 0.935                 & 0.967                  & 0.978                  & 0.221                   & 0.268                   & 0.311                 & 0.383                 & 0.431                  & 0.432                   \\ 
\hline
\multirow{4}{*}{LogAnomaly}              & P  & 0.975                   & 0.656                   & 0.732                 & 0.896                 & 0.950                  & 0.972                  & 0.146                   & 0.183                   & 0.213                 & 0.274                 & 0.321                  & 0.338                   \\
                                         & R  & 0.485                   & 0.905                   & 0.946                 & 0.993                 & 0.994                  & 0.993                  & 0.987                   & 0.979                   & 0.957                 & 0.961                 & 0.971                  & 0.924                   \\
                                         & S  & 1.000                   & 0.998                   & 0.996                 & 0.994                 & 0.995                  & 0.995                  & 0.523                   & 0.521                   & 0.530                 & 0.534                 & 0.552                  & 0.593                   \\
                                         & F1 & 0.647                   & 0.760                   & 0.825                 & 0.942                 & 0.971                  & 0.983                  & 0.255                   & 0.308                   & 0.348                 & 0.427                 & 0.482                  & 0.495                   \\ 
\hline
\multirow{4}{*}{PLELog}                  & P  & 0.947                   & 0.910                   & 0.934                 & 0.972                 & 0.974                  & 0.974                  & 0.583                   & 0.564                   & 0.617                 & 0.765                 & 0.786                  & 0.895                   \\
                                         & R  & 0.972                   & 0.690                   & 0.794                 & 0.956                 & 1.000                  & 1.000                  & 0.691                   & 0.587                   & 0.706                 & 0.872                 & 0.883                  & 0.817                   \\
                                         & S  & 1.000                   & 0.999                   & 0.999                 & 0.999                 & 0.996                  & 0.996                  & 0.915                   & 0.950                   & 0.919                 & 0.942                 & 0.946                  & 0.987                   \\
                                         & F1 & 0.786                   & 0.785                   & 0.859                 & 0.964                 & 0.987                  & 0.987                  & 0.632                   & 0.576                   & 0.659                 & 0.815                 & 0.832                  & 0.855                   \\ 
\hline
\multirow{4}{*}{LogRobust}               & P  & 0.519                   & 0.673                   & 0.794                 & 0.926                 & 0.945                  & 0.964                  & 0.500                   & 0.506                   & 0.632                 & 0.787                 & 0.851                  & 0.819                   \\
                                         & R  & 0.507                   & 0.646                   & 0.999                 & 1.000                 & 1.000                  & 1.000                  & 0.840                   & 0.856                   & 0.896                 & 0.941                 & 0.959                  & 0.924                   \\
                                         & S  & 0.999                   & 0.998                   & 0.997                 & 0.996                 & 0.994                  & 0.994                  & 0.930                   & 0.908                   & 0.969                 & 0.953                 & 0.963                  & 0.954                   \\
                                         & F1 & 0.513                   & 0.659                   & 0.885                 & 0.962                 & 0.972                  & 0.982                  & 0.627                   & 0.636                   & 0.741                 & 0.857                 & 0.902                  & 0.868                   \\ 
\hline
\multirow{4}{*}{CNN}                     & P  & 0.945                   & 0.675                   & 0.793                 & 0.920                 & 0.953                  & 0.967                  & 0.911                   & 0.974                   & 0.901                 & 0.818                 & 0.860                  & 0.865                   \\
                                         & R  & 0.388                   & 0.879                   & 0.947                 & 1.000                 & 0.999                  & 0.999                  & 0.680                   & 0.784                   & 0.791                 & 0.922                 & 0.930                  & 0.901                   \\
                                         & S  & 1.000                   & 0.998                   & 0.997                 & 0.996                 & 0.995                  & 0.995                  & 0.994                   & 0.998                   & 0.988                 & 0.963                 & 0.967                  & 0.968                   \\
                                         & F1 & 0.550                   & 0.763                   & 0.863                 & 0.959                 & 0.976                  & 0.983                  & 0.779                   & 0.869                   & 0.843                 & 0.868                 & 0.894                  & 0.883                   \\
\hline
\end{tabular}
}%
\vspace{-6pt}
\end{table}

Previous studies on software defect prediction models show that prediction results may not always be satisfactory in the presence of imbalanced data distribution~\cite{zhang2007comments}.
Basically, a high probability of detection (i.e., true-positive rate) and a low probability of false alarm (i.e., false-positive rate) do not necessarily lead to high precision due to the imbalanced class distributions. Our results confirm that finding. As a consequence, the commonly used evaluation metrics (Precision, Recall, and F-measure) are not capable of evaluating models in some imbalanced data scenarios and may lead to imprecise evaluation.
Therefore, we propose to used an additional metric, Specificity, to evaluate log-based anomaly detection models more comprehensively. Specificity~\cite{yerushalmy1947statistical, sokolova2006beyond}, which is the percentage of log sequences that are correctly identified normal over all real normal sequences, can measure the probability of false alarms. High Specificity means that models can perform with a low false-positive (false alarms) rate.

\begin{tcolorbox}[left=2pt,right=2pt,top=0pt,bottom=0pt]
\textbf{Summary.}
Highly imbalanced data with a small percentage of anomalies impedes model performance. 
Using different metrics may lead to different conclusions about the model performance. Some commonly-used metrics, including Precision, Recall, and F-measure, are not comprehensive enough for evaluating log-based anomaly detection with highly imbalanced data. More evaluation metrics such as Specificity should be used for a thorough evaluation. 
\end{tcolorbox}


\subsection{RQ4: Can existing methods work with different degrees of data noise?}

\subsubsection{The impact of mislabelled logs}
\label{sec:mislabelled_logs}
\begin{figure}[h]
    \centering
    \includegraphics[width=\linewidth]{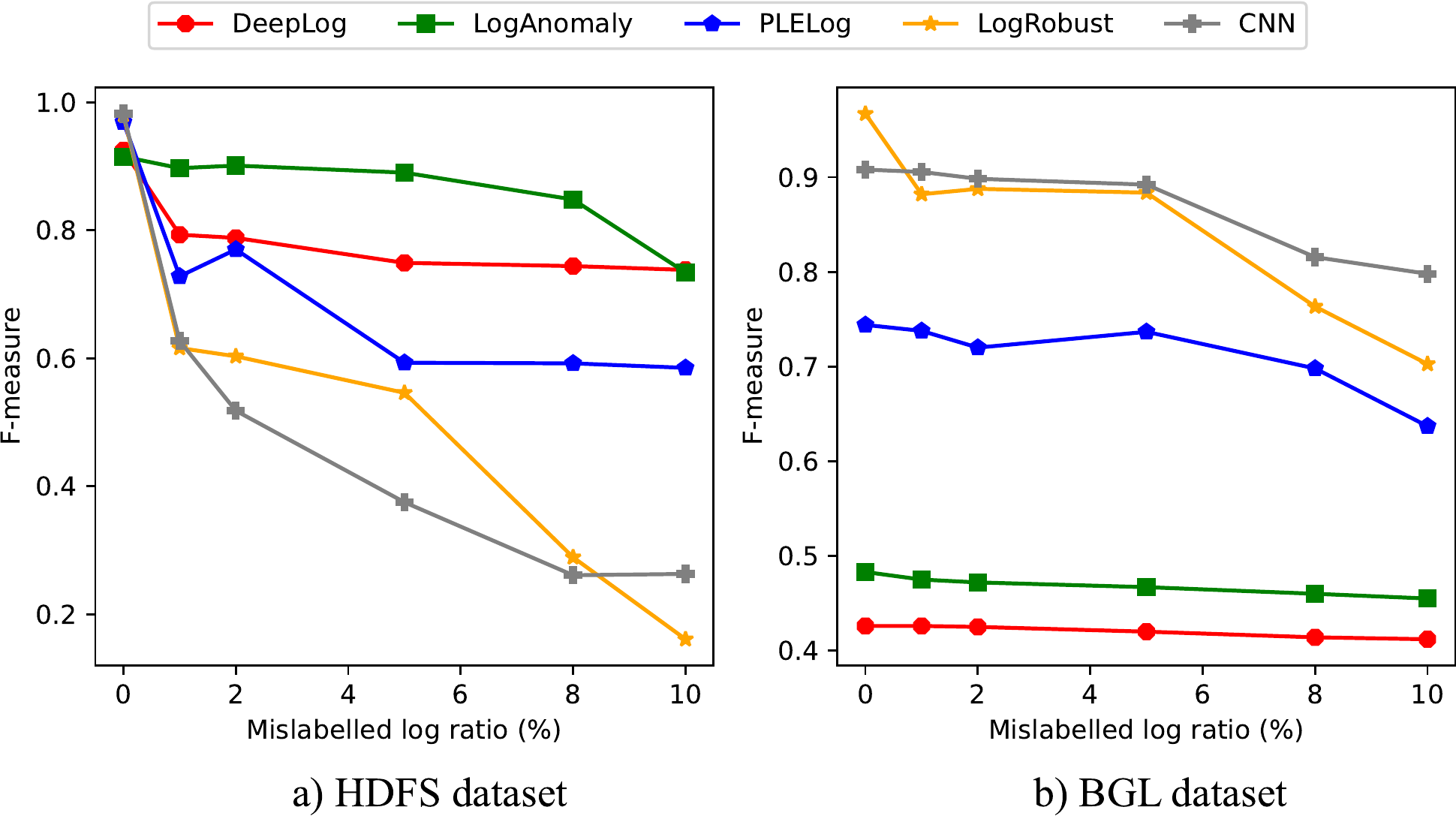}
    \vspace{-6pt}
    \caption{Results with different ratios of mislabelled logs}
    \label{fig:RQ4_mislabeled}
    \vspace{-3pt}
\end{figure}

We evaluate the studied models with different degrees of mislabelled logs.  
In this experiment, we inject a specific portion of mislabelled logs into the training data while the testing sets remain the same. Specifically, for semi-supervised methods, which only use normal logs for training, we put back some anomalies to the training sets. For supervised methods, we change the label of some anomalies in the training sets to normal. \rv{We experiment on all four datasets and find that the performance of models can greatly decrease if training data contains mislabelled samples. We show the results on HDFS and BGL datasets in Figure \ref{fig:RQ4_mislabeled} to demonstrate our finding}.

We can find that, on the HDFS dataset, even with just 1\% of mislabelled logs, the F-measures of the studied models decline significantly. For example, LogRobust and CNN drop 37.2\% and 36.2\% with only 1\% of noise, respectively. When the noise ratio reaches 10\%, the F-measures of LogRobust and CNN drop to only 0.161 and 0.263. The results confirm that even with the advantage of 
semantic understanding, these DL models can lose their performance with only a small proportion of mislabelled logs. It is also true for PLELog, which leverages the semantic meaning of logs as well.
\rv{Another reason for this remarkable reduction is that there are many duplicate log sequences in the HDFS dataset, thus, labeling any duplicate normal sequences as anomalies could have a large impact}.
Interestingly, the F-measure of LogAnomaly only drops slightly with 1\% of mislabelled logs (from 0.915 to 0.897). When the mislabelled ratio is 5\% and 10\%, LogAnomaly can achieve better F-measures compared to other models. The reason is that LogAnomaly uses quantitative vectors to extract quantitative relationships holding in logs along with sequential vectors, thus, allowing the model to predict the possibilities of the next event more precisely.


\begin{tcolorbox}[left=2pt,right=2pt,top=0pt,bottom=0pt]
\textbf{Summary.}
A small amount of mislabelled logs can quickly downgrade the performance of anomaly detection. 
Supervised models are more sensitive to mislabelled logs.
Models adopting the forecasting-based approach (DeepLog and LogAnomaly) perform better with the presence of mislabelled logs.
\end{tcolorbox}

\subsubsection{The impact of log parsing errors}
\label{sec:log_parsing_errors}
We also evaluate the impact of data noise introduced by log parsing errors. Log parsing errors can lead to extra log events and wrong log templates~\cite{le2021log}. We experiment with four commonly-used log parsers, namely Drain~\cite{he2017drain}, Spell~\cite{du2016spell}, AEL~\cite{jiang2008abstracting}, and IPLoM~\cite{makanju2009clustering} \nrv{on all four datasets. We find that the performance of models is highly influenced by log parsers.}
\rv{To demonstrate our findings, we show the performance of models with different log parsers on BGL and Spirit datasets in Table \ref{tab:RQ3.2_noise_log_parser}.}

\begin{table}[h]
\centering
\caption{\rv{Results with different log parsers}}
\vspace{-6pt}
\label{tab:RQ3.2_noise_log_parser}
\resizebox{\linewidth}{!}{
\setlength{\tabcolsep}{2pt}
\renewcommand{\arraystretch}{1.05}
\begin{tabular}{cccccccccc} 
\hline
\multirow{2}{*}{\begin{tabular}[c]{@{}c@{}}\\\textbf{Model }\end{tabular}} &    & \multicolumn{4}{c}{\textbf{BGL}}                                                                   & \multicolumn{4}{c}{\textbf{Spirit}}                                                                 \\ 
\cline{2-10}
                                                                          &    & \textbf{\textit{Drain}} & \textbf{\textit{Spell}} & \textbf{\textit{AEL}} & \textbf{\textit{IPLoM}} & \textbf{\textit{Drain}} & \textbf{\textit{Spell}} & \textbf{\textit{AEL}} & \textbf{\textit{IPLoM}}  \\ 
\hline
\multirow{4}{*}{DeepLog}                                                   & P  & 0.270                   & 0.271                   & 0.271                 & 0.273                   & 0.438                   & 0.602                   & 0.413                 & 0.607                    \\
                                                                          & R  & 0.988                   & 0.988                   & 0.988                 & 0.988                   & 1.0                     & 1.0                     & 1.0                   & 1.0                      \\
                                                                          & S  & 0.437                   & 0.44                    & 0.438                 & 0.443                   & 0.099                   & 0.535                   & 0                     & 0.545                    \\
                                                                          & F1 & 0.426                   & 0.426                   & 0.425                 & 0.427                   & 0.609                   & 0.751                   & 0.584                 & 0.755                    \\ 
\hline
\multirow{4}{*}{LogAnomaly}                                                & P  & 0.313                   & 0.339                   & 0.539                 & 0.548                   & 0.438                   & 0.612                   & 0.413                 & 0.607                    \\
                                                                          & R  & 0.798                   & 0.977                   & 0.965                 & 0.971                   & 1.0                     & 1.0                     & 1.0                   & 1.0                      \\
                                                                          & S  & 0.551                   & 0.599                   & 0.826                 & 0.831                   & 0.099                   & 0.554                   & 0                     & 0.545                    \\
                                                                          & F1 & 0.483                   & 0.504                   & 0.692                 & 0.700                   & 0.609                   & 0.759                   & 0.584                 & 0.755                    \\ 
\hline
\multirow{4}{*}{PLELog}                                                    & P  & 0.702                   & 0.560                   & 0.661                 & 0.655                   & 0.931                   & 0.859                   & 0.863                 & 0.500                    \\
                                                                          & R  & 0.791                   & 0.404                   & 0.854                 & 0.433                   & 0.767                   & 0.859                   & 0.887                 & 0.662                    \\
                                                                          & S  & 0.899                   & 0.938                   & 0.907                 & 0.952                   & 0.690                   & 0.901                   & 0.901                 & 0.535                    \\
                                                                          & F1 & 0.744                   & 0.476                   & 0.744                 & 0.521                   & 0.841                   & 0.859                   & 0.875                 & 0.570                    \\ 
\hline
\multirow{4}{*}{LogRobust}                                                 & P  & 0.994                   & 0.849                   & 0.844                 & 0.726                   & 0.985                   & 0.947                   & 0.986                 & 0.973                    \\
                                                                          & R  & 0.942                   & 0.988                   & 0.982                 & 0.977                   & 0.915                   & 1.0                     & 1.0                   & 1.0                      \\
                                                                          & S  & 0.999                   & 0.963                   & 0.962                 & 0.922                   & 0.990                   & 0.960                   & 0.990                 & 0.980                    \\
                                                                          & F1 & 0.967                   & 0.914                   & 0.908                 & 0.833                   & 0.949                   & 0.973                   & 0.993                 & 0.986                    \\ 
\hline
\multirow{4}{*}{CNN}                                                       & P  & 0.871                   & 0.994                   & 0.942                 & 0.937                   & 0.986                   & 0.959                   & 0.986                 & 0.986                    \\
                                                                          & R  & 0.947                   & 0.965                   & 0.947                 & 0.959                   & 1.0                     & 1.0                     & 1.0                   & 1.0                      \\
                                                                          & S  & 0.97                    & 0.999                   & 0.988                 & 0.986                   & 0.990                   & 0.970                   & 0.990                 & 0.990                    \\
                                                                          & F1 & 0.908                   & 0.979                   & 0.945                 & 0.948                   & 0.993                   & 0.979                   & 0.993                 & 0.993                    \\
\hline
\end{tabular}
}
\vspace{-6pt}
\end{table}

We can observe that the performance of studied models varies a lot with different log parsers. For example, DeepLog achieves an F-measure of 0.755 and a Specificity of 0.545 with the IPLoM~\cite{makanju2009clustering} parser \rv{on Spirit dataset}. When experimenting with Drain~\cite{he2017drain}, these values drop to 0.609 and 0.099, respectively, although Drain is one of the most accurate log parsers according to a recent benchmark study~\cite{zhu2019tools}.
The results also show that LogRobust and CNN can handle log parsing noise better than other models. This is probably because of their use of semantic vectors.
Moreover, we find that different log parsing errors have distinctive impact on the detection models. \rv{The results on BGL and Spirit datasets show that DeepLog and LogAnomaly perform better with Spell~\cite{du2016spell} and IPLoM~\cite{makanju2009clustering} \nrv{than with Drain}. This is because Drain often produces many extra log events that hinder the performance of DeepLog and LogAnomaly (which use forecasting methods to predict the next log event). In contrast, other models can better handle extra log events but may fail when log parsers produce errors due to semantic misunderstanding~\cite{le2021log}.} 

\begin{tcolorbox}[left=2pt,right=2pt,top=0pt,bottom=0pt]
\textbf{Summary.}
The data noise from log parsing errors has impact on the performance of models. Extra log events can quickly downgrade the performance of forecasting-based models. Methods using semantic vectors can better handle log parsing errors. 
\end{tcolorbox}

\subsection{RQ5: How early can the existing models detect anomalies in online detection?}
\label{sec:eval_early_detection}

\begin{figure}[h]
    \centering
    \subfigure[Data grouping with 100 logs]{
    \includegraphics[scale=0.40]{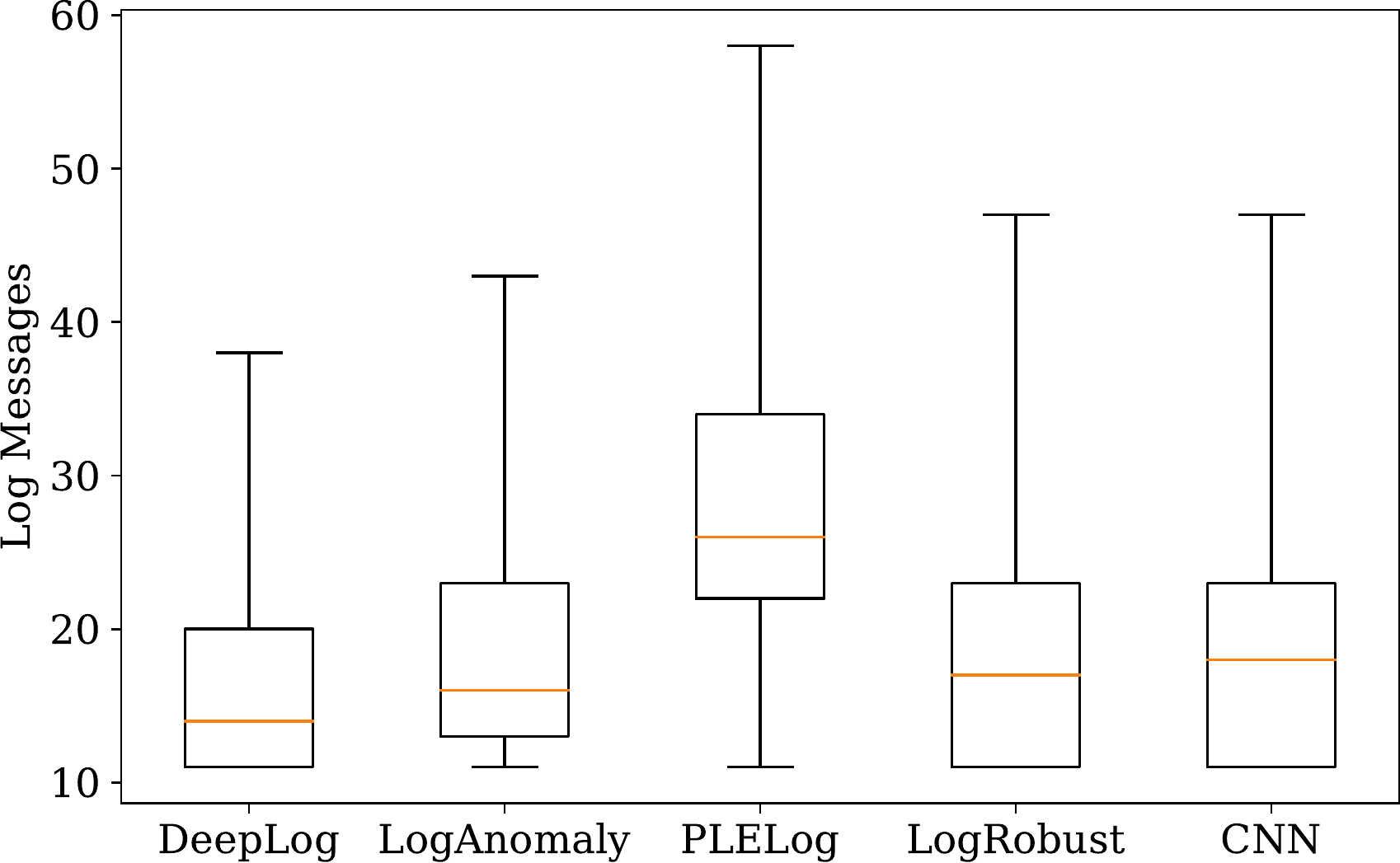}
    \label{fig:100-leadtime}
    }
    \subfigure[Data grouping with 0.5-hour logs]{
    \includegraphics[scale=0.40]{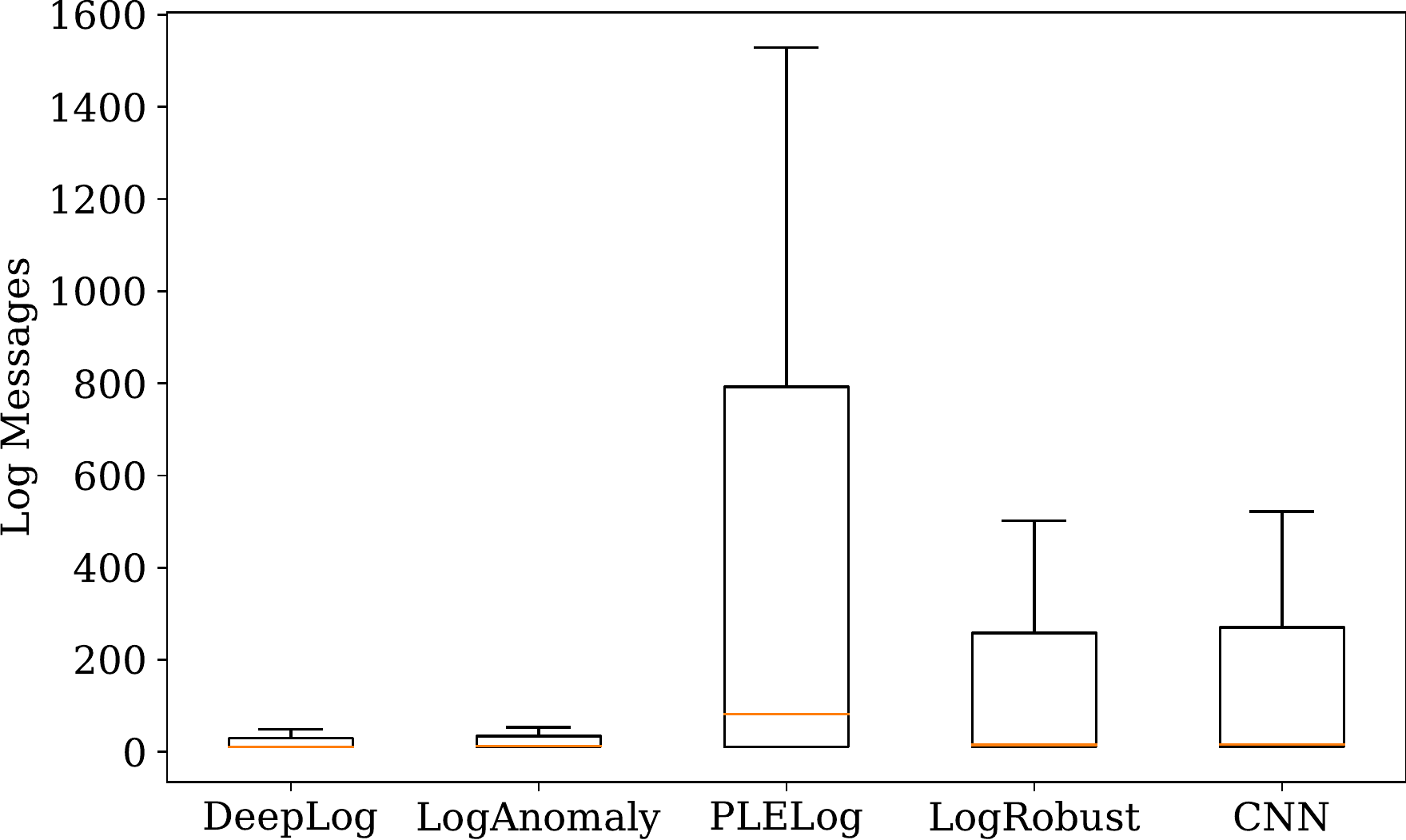}
    \label{fig:0.5h-leadtime}
    }
    \vspace{-6pt}
    \caption{The number of examined log messages before each model can detect an anomaly}
    \label{fig:lead_time}
    \vspace{-6pt}
\end{figure}

To answer RQ5, we evaluate the five studied models using a fixed-window grouping of \rv{different sizes (from 20 logs to 1-hour logs)}. We record the number of examined log messages before each model can detect an anomaly in an online detection mode (i.e., detecting anomalies on the fly). 
Figure \ref{fig:lead_time} uses the results \rv{with 100-logs and 0.5-hour-logs settings to explain our findings}.

\begin{figure}[h]
    \centering
    \includegraphics[width=.97\columnwidth]{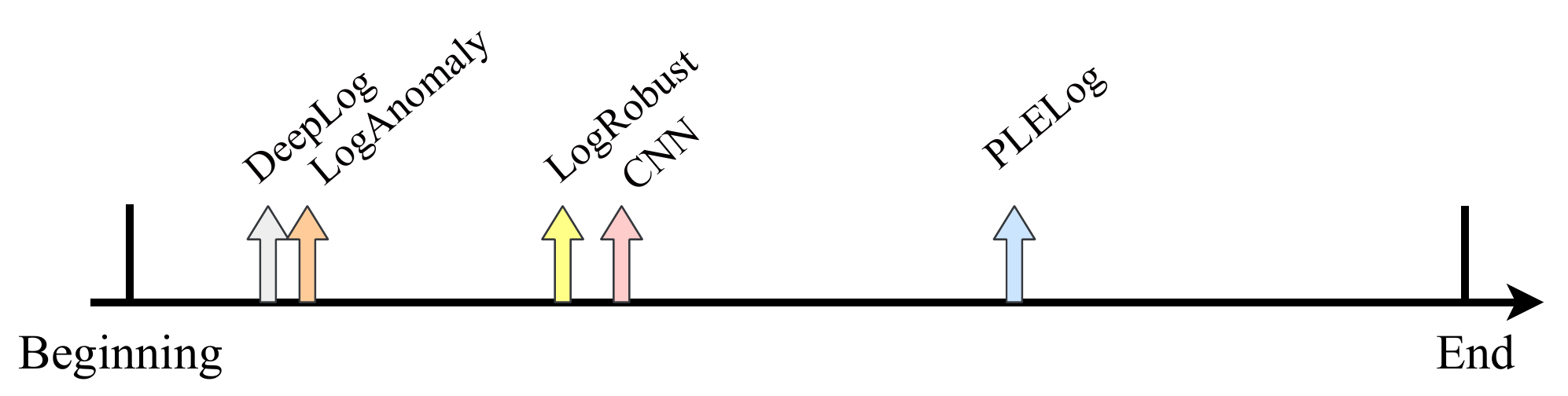}
    \vspace{-6pt}
    \caption{Timeline for detecting an anomaly in the BGL dataset}
    \label{fig:lead_time2}
    \vspace{-6pt}
\end{figure}

We find that DeepLog can detect anomalies the earliest, with an average of \rv{14.3 and 90.6 log messages with 100-logs and 0.5-hour-logs setting, respectively}. LogAnomaly, which is also based on predicting the next log events, detects anomaly a bit later since it requires to capture both sequential and quantitative relationships in order to mark a log sequence as abnormal. 
In contrast, the use of semantic vectors could make other models, including LogRobust, CNN, and PLELog, detect anomalies much later, \nrv{as shown in Figure ~\ref{fig:lead_time}.}
These classification-based models require to capture features of anonymous behavior to detect anomalies, thus tend to raise alarms closer to the time the anomaly happens. Figure~\ref{fig:lead_time2} shows \rv{a case study on an anomalous} log sequence (with the size of 100 logs) on BGL dataset. Five models are applied to identify whether the log sequence is abnormal or not. We find that DeepLog raises an alarm after examining 11 logs, followed by LogAnomaly, LogRobust, and CNN, while PLELog is the last model that raises an alarm (after 64 log messages arrived).

As efficiency is critical in online anomaly detection, we also evaluate the studied models by recording the time spent on both the training and testing phases. On BGL dataset, DeepLog spends 19.2 and 0.6 minutes on training and testing. LogRobust and CNN, which use higher-dimension inputs and more complex networks, consume 31.7 and 28.3 minutes on training and testing, respectively. In the testing phase, LogRobust and CNN only take 0.3 minutes. LogAnomaly, which uses two LSTM networks to learn sequential and quantitative features, spends 56.7 and 1.6 minutes on training and testing. PLELog, which contains a clustering module and a GRU module, is the most time-consuming model. Specifically, PLELog spends 36.6 minutes and 69.1 minutes to train the clustering and GRU models, respectively. In the testing phase, PLELog consumes 10.8 minutes to process the BGL dataset. The results show that PLELog may be inappropriate for online detection due to its incapability of early detection and heavy time cost.

\begin{tcolorbox}[left=2pt,right=2pt,top=0pt,bottom=0pt]
\textbf{Summary.}
Different models have different abilities in the early detection of system anomalies. Forecasting-based models (DeepLog and LogAnomaly) can detect anomalies earlier than classification-based models (PLELog, LogRobust, and CNN).
\end{tcolorbox}

\vspace{-1mm}
\section{Discussion}
\label{sec:discussion}
\subsection{The Advantages and Disadvantages of the Studied Methods}

Based on our findings, we can conclude that all the studied models do not always work as well as they claimed in their papers. Different scenarios have different impacts on the performance of anomaly detection models.
We point out the advantages and disadvantages of each model as follows, which are also summarized in Table \ref{tab:summary_models}.

\textbf{DeepLog}. 
The main advantage of DeepLog is that it does not require any abnormal logs to build a detection model using sequential vectors, thus reducing the effort for model construction. 
It can also detect anomalies earlier than other models.
However, due to this characteristic, DeepLog performs poorly on more complex datasets with a large number of log events (see Section \ref{sec:eval_training_data}). \nrv{Besides, DeepLog is greatly impacted by the log parsing errors since it only leverages the index of log templates and ignores the semantic meaning of log templates (see Section \ref{sec:log_parsing_errors}).}

\textbf{LogAnomaly}.
LogAnomaly uses sequential and quantitative vectors to train a detection model, which can help reduce the impact of data noise caused by mislabelled logs.
Like DeepLog, LogAnomaly can detect anomalies early and deal with a large amount of data since it only trains with normal logs. The main benefit of LogAnomaly is in the phase of matching similar log templates using semantic vectors. This feature allows LogAnomaly to improve the accuracy by matching new log templates with an existing one in the training logs instead of marking them as anomalies as DeepLog does. However, like DeepLog, LogAnomaly trains the model using the index of log event (sequential and quantitative vectors) \nrv{and}  cannot learn the semantic meaning of log templates, so it is highly affected by the log parsing errors (see Section \ref{sec:log_parsing_errors}). Moreover, LogAnomaly cannot perform well on large datasets with a numerous log events (see Section \ref{sec:eval_training_data} and \ref{sec:eval_data_grouping}).

\textbf{PLELog}.
The main advantage of PLELog is that it can learn knowledge about historical anomalies via probabilistic label estimation. 
PLELog adopts a clustering method (i.e., HDBSCAN~\cite{mcinnes2017hdbscan}) to probabilistically estimate the labels of unlabeled log sequences. This approach allows PLELog to work 
with only normal logs. Besides, the use of semantic vectors and attention-based GRU network makes PLELog perform more effectively. However, PLELog is time-consuming since it requires time to train the clustering model (see Section \ref{sec:eval_early_detection}).
Moreover, PLELog cannot cope well with the noise in training data (see Section \ref{sec:mislabelled_logs}), and it does not perform well on the early detection task.

\textbf{LogRobust}.
LogRobust leverages semantic vectors of log templates together with an attention-based Bi-LSTM model. LogRobust can work well with the noise from log parsing errors by utilizing the attention-based Bi-LSTM model, which has the ability to capture the contextual information of log sequences. 
However, as the main characteristic of supervised models, LogRobust requires both normal and abnormal data in the training phase, which would cost much manual labeling effort. Another drawback of LogRobust is that it can be greatly affected by the noise from mislabelled logs (see Section \ref{sec:mislabelled_logs}). 

\textbf{CNN}.
The Convolutional Neural Network can mine more relationships in log context by leveraging multiple filters. The convolution operation allows CNN to capture not only the correlation between log templates but also the correlation inside the semantic embedding of log templates~\cite{lu2018detecting}. CNN can achieve high accuracy on many datasets using a supervised approach. However, like LogRobust, CNN requires a large amount of labeled data, which is mostly unavailable in practice. CNN also loses its accuracy when dealing with highly imbalanced data and data noise (see Sections \ref{sec:eval_data_distribution} and \ref{sec:mislabelled_logs}).


\begin{table*}[htbp]
\centering
\caption{A Comparison of Different Log-based Anomaly Detection Approaches with Deep Learning}
\vspace{-6pt}
\label{tab:summary_models}
\resizebox{.96\linewidth}{!}{%
\setlength{\tabcolsep}{3pt}
\renewcommand{\arraystretch}{1.18}
\begin{tabular}{|c|L{6.8cm}|L{6.5cm}|L{3cm}|} 
\hline
\textbf{Model} & \multicolumn{1}{c|}{\textbf{Pros}}                                                                                                                    & \multicolumn{1}{c|}{\textbf{Cons}}                                                                             & \multicolumn{1}{c|}{\textbf{Data Requirements}}  \\ 
\hline
DeepLog        & Relatively simple, only require normal data. Good at detecting anomalies early detection.                                                             & Perform poorly on complex datasets. Heavily impacted by the log parsing errors.                                & Normal labeled and simple data.                    \\ 
\hline
LogAnomaly     & Only require normal data. Can match new log templates with existing ones in the training data.                                                        & Perform poorly on complex datasets. Heavily impacted by the log parsing errors.                  
& Normal labeled and simple data.                    \\ 
\hline
PLELog         & Only require normal and unlabelled data. Good performance compared to semi-supervised methods. Can learn the semantic meaning of log templates. & Cannot work well with the noise in training data and in the early detection task. Complex design and time-consuming. & Partially labeled data.                             \\ 
\hline
LogRobust      & Good performance on many datasets. Can capture the contextual information of log sequences to handle log parsing errors and the instability of logs.          & Require a large amount of labeled data. Heavily impacted by mislabelled logs.            & Fully labeled data.                                \\ 
\hline
CNN            & Good performance on many datasets. 
Can handle log parsing errors.  
& Require a large amount of labeled data. Heavily impacted by mislabelled logs.            & Fully labeled data.                                \\
\hline
\end{tabular}
}
\end{table*}

\vspace{-6pt}
\subsection{Future Research Work}

Based on our findings, we identify the following research challenges and also suggest possible solutions:

\textit{A variety of datasets.} Our findings suggest that more datasets should be used for a more comprehensive evaluation of log-based anomaly detection models. A good result on one dataset does not necessarily reflect good performance on other datasets due to a variety of data characteristics (e.g., class distributions, noise, etc.).

\textit{Limited labelled data.} Although supervised learning-based methods (i.e., LogRobust and CNN) can achieve higher accuracy than the unsupervised counterparts, it is time-consuming and tedious to manually label the anomalies due to the volume and velocity of log data. Semi-supervised learning-based models (i.e., DeepLog, LogAnomaly, and PLELog) can deal with a large amount of data as they only require normal data. However, the accuracy achieved by existing methods is rather low in practice, as shown in Section \ref{sec:eval_training_data}. Improving the accuracy of semi-supervised models or designing unsupervised models is a challenging but essential future work. 
    
\textit{Early detection.}
Our findings show that different models have different abilities in early detection of system anomalies. The anomalies should be predicted as early as possible to allow enough time for any preparatory or preventive actions in an online detection scenario. More work is needed to build effective models that allow sufficient lead time (i.e., reducing the number of log messages being examined) and meanwhile achieve high prediction accuracy.
    
\textit{Evolving systems.}
Our findings show that log parsing errors have impact on log-based anomaly detection. Log parsing errors can be introduced by the change of logging statements during software evolution~\cite{zhang2019robust, le2021log}. 
As real-world software systems constantly evolve, new log events always appear~\cite{zhang2019robust, kabinna2018examining}. Therefore, existing log-based anomaly detection models will either fail to work due to the incompatibility with new logs or result in low performance due to the incorrect classification. 
Therefore, models should be able to learn the semantic meaning of the whole log messages to handle the instability of logs of evolving systems.


\textit{Relations among log events.}
As discussed in Section \ref{sec:eval_data_grouping}, session window grouping gathers logs in a specific execution path, which could possess more relations among log events than the fixed-window grouping method. Existing methods convert logs into sequences, which capture sequential relationships among log messages. A possible research direction is to explore more relationships between logs, such as the logical relationships and interactive relationships among logging components~\cite{liu2019log2vec}, to capture a variety of anomalous behavior.

\vspace{-6pt}
\subsection{Threats to Validity}

During our study, we have identified the following major threats to the validity.

\textbf{Limited models}. In this work, we only experimentally evaluate five representative models that have publicly available source code. In the future, we will aim to re-implement the DL models that did not release their source code based on the descriptions in their papers and then perform a larger-scale evaluation.

\rv{\textbf{Reimplementation}. We mainly adopt the public implementations of studied models. For LogRobust, as its original implementation is based on Keras, we convert it into a PyTorch-based implementation so that we have a unified framework for all log-based anomaly detection tools. In our version, we use the same hyperparameters that are provided by the authors of LogRobust. For LogAnomaly and CNN, we use the FastText model to replace the missing semantic embedding components from public implementations. To reduce this threat, we experiment on the same settings and datasets from the original paper and confirm that our results are similar with the reported values.}

\textbf{Limited datasets}. Our experiments are conducted on four public log datasets. Although they are widely used in existing studies on log-based anomaly detection, they may not represent all characteristics of log data. To overcome this threat, we create synthetic datasets for evaluating models with different data characteristics (e.g., different class distributions and different labeling noise). In the future, we will experiment on more datasets, including industrial datasets, to cover more real-world scenarios.

\textbf{Data quality}. Our experiments are conducted based on four datasets that are manually inspected and labeled by engineers.
However, our experiment on data noises shows that a small portion of mislabelled logs could downgrade the performance of anomaly detection models. To reduce this threat, we experiment with four public datasets. We also create synthetic datasets by injecting a specific portion of mislabelled logs. In our future work, we will
explore methods for measuring and improving the quality of datasets. 


\section{Conclusion}
We have conducted an in-depth analysis of recent deep learning models for log-based anomaly detection. We have investigated several aspects of model evaluation: training data selection strategies, different characteristics of datasets, and early detection capability. Our results point out that all these aspects have large impact on the evaluation results \rv{and the performance of the models is often not as good as expected. Our findings show that the problem of log-based anomaly detection has not been solved yet.}
We also suggest some possible future work. 
We hope that the results and findings of our study can be of great help for practitioners and researchers working on this interesting area.

Our source code and detailed experimental data are available at \textbf{\url{https://github.com/LogIntelligence/LogADEmpirical}}. \rv{The datasets including the synthetic data that is used for evaluating models with different data characteristics, can also serve as a benchmark for evaluating future log-based anomaly detection models.}

\begin{acks}
This work is supported by Australian Research Council (ARC) Discovery Projects (DP200102940, DP220103044).
We also thank anonymous reviewers for their insightful and constructive comments, which significantly improve this paper.

\end{acks}

\balance
\bibliographystyle{ACM-Reference-Format}
\bibliography{sample}
\balance

\end{document}